\documentclass[]{article}

\usepackage{graphicx}
\usepackage[capposition=top]{floatrow}
\usepackage[margin=1.in]{geometry}
\usepackage{amsmath}
\usepackage{amssymb}
\usepackage{float}
\usepackage{hyperref}
\usepackage{setspace}
\usepackage{multicol,lipsum}
\usepackage{algorithm}
\usepackage{algpseudocode}

\newcommand{\edit}[1]{{\color{black}#1}}

\title{Empirical sandwich variance estimator for iterated conditional expectation g-computation}
\author{Paul N Zivich\textsuperscript{1,2}, Rachael K Ross\textsuperscript{2}, Bonnie E Shook-Sa\textsuperscript{3}, Stephen R Cole\textsuperscript{2}, Jessie K Edwards\textsuperscript{2}}
\date{%
	\small	
	\textsuperscript{1}Institute of Global Health and Infectious Diseases, University of North Carolina at Chapel Hill, Chapel Hill, NC\\%
	\textsuperscript{2}Department of Epidemiology, Gillings School of Global Public Health, University of North Carolina at Chapel Hill, Chapel Hill, NC\\%
	\textsuperscript{3}Department of Biostatistics, Gillings School of Global Public Health, University of North Carolina at Chapel Hill, Chapel Hill, NC\\%
	\today
}

\begin{document}
	
\maketitle

\begin{abstract}
	Iterated conditional expectation (ICE) g-computation is an estimation approach for addressing time-varying confounding for both longitudinal and time-to-event data. Unlike other g-computation implementations, ICE avoids the need to specify models for each time-varying covariate. For variance estimation, previous work has suggested the bootstrap. However, bootstrapping can be computationally intense. Here, we present ICE g-computation as a set of stacked estimating equations. Therefore, the variance for the ICE g-computation estimator can be consistently estimated using the empirical sandwich variance estimator. Performance of the variance estimator was evaluated empirically with a simulation study. The proposed approach is also demonstrated with an illustrative example on the effect of cigarette smoking on the prevalence of hypertension. In the simulation study, the empirical sandwich variance estimator appropriately estimated the variance. When comparing runtimes between the sandwich variance estimator and the bootstrap for the applied example, the sandwich estimator was substantially faster, even when bootstraps were run in parallel. The empirical sandwich variance estimator is a viable option for variance estimation with ICE g-computation.
\end{abstract}

\section{Introduction}

Causal inference with longitudinal and time-to-event data often must contend with time-varying confounding, whereby a covariate is both a confounding variable and is affected by prior treatment \cite{robins_marginal_2000}. One approach to appropriately address time-varying confounding is the g-formula \cite{robins_new_1986}. Two g-formula estimators are standard g-computation \cite{robins_new_1986, taubman_intervening_2009, keil_parametric_2014, westreich_parametric_2012}, and iterated conditional expectation (ICE) g-computation \cite{kreif_estimating_2017, schomaker_using_2019, wen_parametric_2021, wen_multiply_2022, rudolph_simulation_2023}. To apply standard g-computation, one specifies models for the outcome and each time-varying covariate. The requirement to specify a model for \textit{each} time-varying covariate has led to concerns regarding correct specification of each of these models. ICE g-computation has the advantage of only requiring specification of outcome models \cite{wen_parametric_2021, rudolph_simulation_2023}.

To consistently estimate the variance for ICE g-computation, previous work has suggested the nonparametric bootstrap \cite{schomaker_using_2019, wen_parametric_2021, rudolph_simulation_2023}. However, the bootstrap is computationally demanding, as it requires repeating the analysis using resamples of the data \cite{efron_bootstrap_1979, kulesa_sampling_2015}. This computational complexity can limit the scenarios considered by researchers in practical applications (e.g., exploring alternative treatment plans, varying functional form specifications, sensitivity analyses). The computational complexity also makes simulation experiments difficult. For example, some simulation studies forgo estimation of the variance by bootstrap in each iteration \cite{wen_parametric_2021, wen_multiply_2022}. While other simulation studies have used the bootstrap \cite{rudolph_simulation_2023}, the computational burden may limit the sample sizes, number of iterations, or scenarios considered. An alternative variance estimator is based on the influence curve, but this estimator is not consistent for ICE g-computation \cite{schomaker_using_2019}.

Here, the ICE g-computation estimator is expressed as a set of estimating equations \cite{stefanski_calculus_2002}, which allows the asymptotic variance of the ICE estimator to be consistently estimated using the empirical sandwich variance estimator \cite{wen_parametric_2021}. The empirical sandwich variance estimator has also been used for variance estimation with g-computation in settings without time-varying confounding \cite{stephens_augmented_2012, sjolander_regression_2016, reifeis_assessing_2020, tackney_comparison_2023}. Here, we illustrate the use of \edit{estimating equations} for ICE g-computation in settings with time-varying confounding. The proposed approach has the usual benefits of ICE g-computation, in addition to providing a statistically consistent variance estimator that is more computationally efficient than the nonparametric bootstrap. This approach also allows one to easily estimate the variance for transformations of parameters and incorporate additional nuisance models by stacking estimating equations together. Finally, the ICE g-computation estimator can be readily implemented using existing software for generic \edit{estimating equations} \cite{zivich_delicatessen_2022, saul_calculus_2020}.

The structure of the paper is as follows. In section 2, the data and a sufficient set of identification assumptions for longitudinal data structures are reviewed. Section 3 reviews g-computation estimators in the setting of repeated measures and presents ICE g-computation as stacked estimating equations. The proposed \edit{implementation} is examined with a simulation study in section 4. In section 5, the proposed ICE g-computation procedure is demonstrated in an illustrative example of estimating the effect of cigarette smoking on prevalent hypertension. Finally, section 6 summarizes the key results and notes how an estimating equation approach for causal effect estimation with longitudinal data can be expanded upon in future work.

\section{Observed data and identification}

Let $k \in \{1, ..., \tau\}$ index discrete follow-up times. The potential outcome at $\tau$ for unit $i$ under the treatment plan (i.e., a defined sequence of treatments) $\bar{a}_{\tau-1}^{*} = (a_0^*, a_1^*, ..., a_{\tau-1}^*)$ is denoted by $Y_{i, \tau}(\bar{a}_{\tau-1}^{*})$. Only deterministic plans of binary treatments (i.e., units are assigned to specific treatments, as opposed to probabilities of treatments) are considered hereafter. A simple example is always treat, where $\bar{a}_{\tau-1}^{*} = (1, 1, ..., 1)$. The parameter of interest is the mean potential outcome at $\tau$ under a given plan (i.e., the joint effect of treatment), $\mu_\tau(\bar{a}_{\tau-1}^{*}) = \mathbb{E}[Y_{i, \tau}(\bar{a}_{\tau-1}^{*})]$, where $\mathbb{E}[\cdot]$ is the expected value function.

For each unit $i$ at time $k$, the observed data consists of the treatment ($A_{i,k}$), a set of covariates ($L_{i,k}$), a loss to follow-up indicator (i.e., censoring status, $C_{i,k}$), and the observed outcome for those uncensored ($Y_{i,k}$). Data are assumed to occur in a specific time-order, namely $L_0 \rightarrow A_0 \rightarrow C_1 \rightarrow Y_1 \rightarrow L_1 \rightarrow A_1 \rightarrow ... \rightarrow Y_\tau$. Here, $Y_{i,k}$ is measured regardless of $Y_{i, k-1}$ (i.e., repeated measures). Overbars are used to indicate the history of a variable, e.g., $\bar{A}_{i,k} = (A_{i,0}, A_{i,1}, ..., A_{i,k})$. Lastly, units lost to follow-up are unobserved for all following time points (i.e., loss to follow-up is monotonic). The observed data consists of $n$ iid units of $O_i = (L_{i, 0}, A_{i,0}, C_{i,1}, Y_{i,1}, L_{i,1}, A_{i,1}, ..., Y_{i,\tau})$ from a random sample of the target population.

Inference for $\mu_\tau(\bar{a}_{\tau-1}^{*})$ in observational studies, or whenever $A_k$ is not randomly assigned according to a known mechanism, is generally complicated by time-varying confounding. Consider the causal diagram in Figure 1. Here, $L_k$ is a time-varying confounder as it is both (1) affected by prior values of treatment, and (2) affects later values of treatment and outcomes. In this setting, $L_k$ functions as a confounding variable for $A_k$ and as a mediating variable for $A_{k-1}$. So, the joint effect of time-varying treatments can no longer be estimated with standard regression methods regardless of whether one adjusts for $L_k$ or not. This can be seen by noting that adjusting for $L_k$ blocks part of the effect of $A_{k-1}$ and not adjusting for $L_k$ means there is still confounding for $A_k$. This concern also extends to cases where prior outcomes determine future treatment, such that there is feedback loop from $A_{k-1}$ to $Y_k$ to $A_{k}$. Hereafter, we let the set of covariates $L_k$ possibly include $Y_k$. For identification and estimation in the cases of time-varying confounding, one can instead use Robins's g-methods, which include the g-formula \cite{robins_new_1986}, inverse probability weighting of marginal structural models \cite{robins_marginal_2000}, and g-estimation of structural nested models \cite{robins_g_1992}. Importantly, each of the g-methods are capable of estimating the joint effect of treatments with time-varying confounding under confounding structures like those depicted in Figure 1.

\begin{figure}[h]
	\centering
	\caption {Causal diagram depicting time-varying confounding by $L_k$}
	\includegraphics[width=0.475\linewidth]{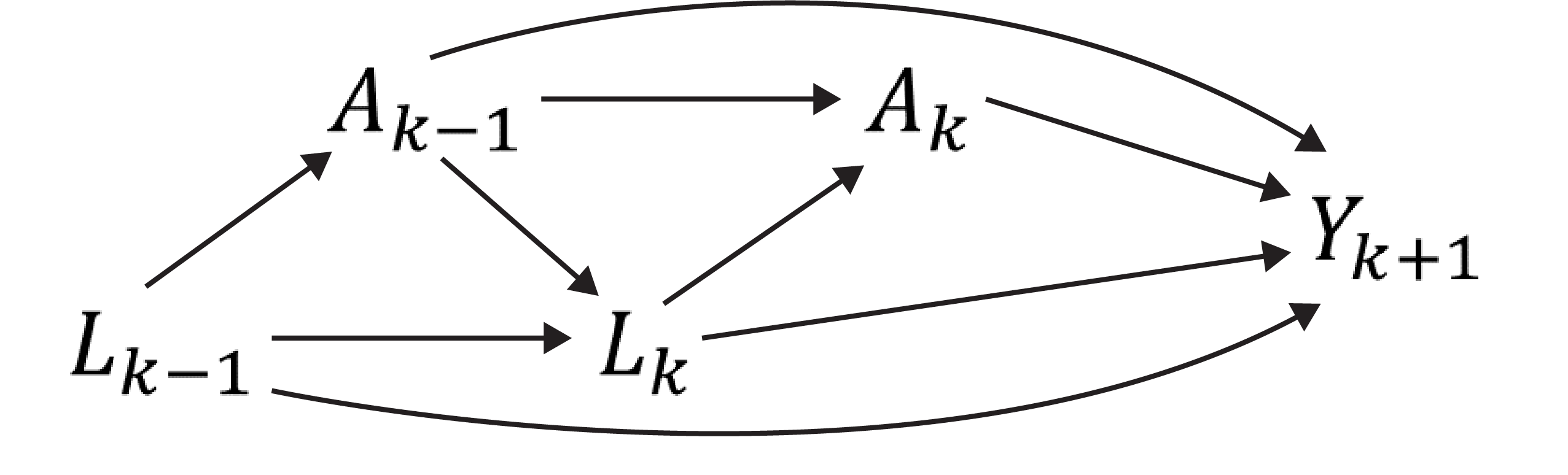}
\end{figure}

To identify $\mu_\tau(\bar{a}_{\tau-1}^{*})$, we proceed following the assumptions provided in Table 1. Comprehensive discussion of these assumptions can be found elsewhere \cite{keil_parametric_2014, wen_parametric_2021, robins_advances_2009, young_comparative_2011, daniel_methods_2013, bodnar_marginal_2004, hernan_chapter_2023}. Briefly, causal consistency provides a connection between the observed outcomes and time-varying covariates with the potential outcomes and covariates under the treatment plans. Causal consistency, as expressed here, implies both no interference between units and variations of treatment are irrelevant \cite{vanderweele_concerning_2009}. Treatment exchangeability stipulates that the treatment at time $k$ is independent of the potential outcomes conditional on the history of previous treatment and covariates. In other words, $L$ includes all the confounding variables. In the context of deterministic plans, treatment positivity states that there is a non-zero probability of following the plan up to $k$ conditional on having followed the plan up to $k-1$ and time-varying confounding history that occur. Censoring exchangeability specifies that loss to follow-up at time $k$ is non-informative conditional on $\bar{L}_k,\bar{A}_k, \bar{Y}_k$. Again, censoring positivity ensures that the expression for censoring exchangeability is well-defined. In addition to these assumptions, we further assume no measurement error. Following these additional assumptions, the parameter of interest can be written as
\begin{equation}
	\mu_\tau(\bar{a}_{\tau-1}^{*}) = \int_{\bar{l}_{\tau-1} \in \bar{L}_{\tau-1}} 
	\left[\mathbb{E}\left\{ Y_\tau | \bar{A}_{\tau-1} = \bar{a}_{\tau-1}^*, \bar{L}_{\tau-1} = \bar{l}_{\tau-1} \right\}
	\prod_{k=1}^{\tau-1} f_{\bar{l}}\left(\bar{l}_{k} | \bar{A}_{k-1} = \bar{a}_{k-1}^*, \bar{L}_{k-1} = \bar{l}_{k-1}\right)\right] 
	d\bar{l}
	\label{equation1}
\end{equation}
where $f_{\bar{l}}(.)$ is the probability density function for $\bar{l}_k$ \cite{robins_new_1986}. Note these expressions are also conditional on having not been lost to follow-up at the prior time (i.e., $C=0$), but this is left implicit for notational simplicity. As the parameter of interest is expressed in terms of observables (i.e., identified), one can consider estimators based on this expression.

\begin{table}[]
	\caption{Sufficient identification assumptions for the mean at the end of follow-up with time-varying confounding and informative censoring}
	\centering
	\setstretch{2.}
	\begin{tabular}{lcc}
		\hline
		Assumption name           & Assumption expression                                                                                                                                                                                                                                   & Condition\textsuperscript{a}                                                                                         \\ \hline
		Causal Consistency        & 
		\setstretch{1.}
		\begin{tabular}[c]{@{}c@{}}
			$Y_{i,k} = Y_{i,k}(\bar{a}_{k-1}^*)$ \\ 
			$L_{i,k} = L_{i,k}(\bar{a}_{k-1}^*)$
		\end{tabular} 		\setstretch{2.} 
	    & if $\bar{a}_{k-1}^* = \bar{A}_{i,k-1}$                \\
		Treatment exchangeability & $Y_{k}(\bar{a}_{k-1}^*) \amalg \bar{A}_{k-1} | \bar{A}_{k-2} = \bar{a}_{k-2}^*, \bar{L}_{k-1}$ & for $\bar{a}_{k-1}^*$                                                                             \\
		Treatment positivity\textsuperscript{b}      & $ f(\bar{a}_k^* | \bar{a}_{k-1}^*, \bar{l}_{k-1}) > 0 $                                                                                        &
		\setstretch{1.}
		\begin{tabular}[c]{@{}c@{}}
			for $\bar{a}_{k-1}^*$, $\bar{l}_{k-1}$ \\ 
			where $f(\bar{a}_{k-1}^*, \bar{l}_{k-1}) > 0$
		\end{tabular}
		\setstretch{2.}
		\\
		Censoring exchangeability & $Y_{k} \amalg C_{k} | \bar{A}_{k-1}, \bar{L}_{k-1}, C_{k-1} = 0$                                                                                                                     & for $C_{k} = 0$                                                                                   \\
		Censoring positivity\textsuperscript{b}      & $ \Pr(C_{k} = 0 | \bar{a}_{k-1}^*, \bar{l}_{k-1}, C_{k-1} = 0) > 0 $                 & 
		\setstretch{1.}
		\begin{tabular}[c]{@{}c@{}}
			for $\bar{a}_{k-1}^*$, $\bar{l}_{k-1}$ \\ 
			where $f(\bar{a}_{k-1}^*, \bar{l}_{k-1}, C_{k-1} = 0) > 0$
		\end{tabular}
		\\ \hline
	\end{tabular}
	\setstretch{1.}
	\floatfoot{
		Overbars indicated history. $Y_{i,k} (\bar{a}_{k-1}^*)$ and $L_{i,k} (\bar{a}_{k-1}^*)$ are the potential outcomes and potential covariates at time k under the plan, respectively. Here, $\bar{L}_k$ may include $\bar{Y}_k$.\\
		\textsuperscript{a} All assumptions are assumed to hold for all $k \in \{1, ..., \tau\}$.\\
		\textsuperscript{b} $f$ is the probability density function for the corresponding variables.
	}
\end{table}

\section{G-computation}

\subsection{Standard g-computation}

One approach for estimation of $\mu_\tau(\bar{a}_{\tau-1}^{*})$ is the parametric generalized computation algorithm formula where time-varying variables are simulated forward in time using a series of parametric models \cite{robins_new_1986}, which we refer to as `standard' g-computation. Here, we briefly review how standard g-computation with pooled regression models is implemented in order to describe its deficiencies compared to ICE g-computation. More detailed descriptions of how to implement standard g-computation can be found in the following references \cite{westreich_parametric_2012, keil_parametric_2014, schomaker_using_2019, mcgrath_gformula_2020}.

Briefly, one converts their data into a repeated measures structure where rows represent an individual per unit of time, referred to as a `long' data set. In this structure, each individual can contribute multiple rows, but will only contribute one row per unit of time. The long data set is then used to fit pooled regression models for $Y_{k}$ conditional on $\bar{A}_{k-1},\bar{L}_{k-1}$ and time, and each variable in $L_{k}$ conditional on $\bar{A}_{k-1},\bar{L}_{k-1}$ and time \cite{abbott_logistic_1985}. If $L_{k}$ is of dimension $p$, then a total of $p+1$ pooled regression models need to be fit. These estimated models are then used along with a Monte Carlo procedure to the simulate time-varying variables under the plan. To simulate outcomes under the plan, the observed values of $L_{0}$ are used as the starting point. Next, $A_0$ is set according to the plan of interest. Values for $L_{0}$ and $A_0$ are then used to simulate $Y_1$ and $L_1$ using the previously estimated models. The process of setting $A_k$ and simulating $Y_{k+1}$ and $L_{k+1}$ given $(\bar{L}_k, \bar{A}_k)$ is repeated until $\tau$. As this procedure generates a set of complete observations under the joint treatment plan, the mean of the simulated values for $Y_\tau$ can be used to estimate $\mu_\tau(\bar{a}_{\tau-1}^{*})$. For causal contrasts, the preceding process is repeated for the alternative treatment plan and then the pair of point estimates are contrasted.

As standard g-computation requires one to specify models for $Y$ and each $L$, it is considered to be highly susceptible to bias resulting from model misspecification \cite{keil_parametric_2014}. Specifically, the standard g-computation estimator is only consistent for $\mu_\tau(\bar{a}_{\tau-1}^{*})$ when \textit{each} of the time-varying covariate models is correctly specified. This assumption becomes more questionable as the number of time-varying confounding variables, $p$, increases. This is a prominent concern in applied settings, where the number of time-varying covariates is likely to be large. As sufficient information to inform model selection for all time-varying covariates is unlikely in practice, other algorithms for g-computation that do not require one to specify models for $L_k$ have been proposed \cite{wen_parametric_2021}.

\subsection{ICE g-computation}

To avoid specifying models for each time-varying covariate in $L_{k}$, the g-formula in Equation \ref{equation1} can instead be written as a series of nested conditional expectations
\begin{equation}
	\label{equation2}
	\mu_\tau(\bar{a}_{\tau-1}^{*}) = 
	\mathbb{E}\left\{ ... 
		\mathbb{E}\left[ 
			\mathbb{E} (Y_\tau | \bar{A}_{\tau-1} = \bar{a}_{\tau-1}^*, \bar{L}_{\tau-1}) |  
		\bar{A}_{\tau-2} = \bar{a}_{\tau-2}^*, \bar{L}_{\tau-2}\right] ... | 
	\bar{A}_{0} = \bar{a}_{0}^*, \bar{L}_{0}\right\}
\end{equation}
where the inner expectation is the outcome at $\tau$ conditional on the plan and covariate history up to that time \cite{robins_new_1986, robins_causal_1997, wen_parametric_2021}. Again, the conditioning on $C_k=0$ is left implicit for notational simplicity. Equation \ref{equation2} leads to the ICE g-computation estimator, which can be implemented with a series of outcome regression models moving backwards through time. The following algorithm can be used to implement ICE g-computation for repeated measures:
\begin{enumerate}
	\item Fit a regression model for $Y_{i,\tau}$ conditional on $\bar{A}_{i,\tau-1}$ and $\bar{L}_{i,\tau-1}$ for all observations where $C_{i,\tau} = 0$.
	\item Generate predicted values of $Y_\tau$ for the plan of interest, $\bar{a}_{i,\tau-1}^*$, and the observed $\bar{L}_{i,\tau-1}$ for all units uncensored at $\tau-1$ (i.e., $C_{i,\tau-1} = 0$). Let the predictions be denoted by $\tilde{Y}_{i,\tau-1}^*$.
	\item Fit a regression model for $\tilde{Y}_{i,\tau-1}^*$ conditional on $\bar{A}_{i,\tau-2}$ and $\bar{L}_{i,\tau-2}$ with all observations where $C_{i,\tau-1} = 0$.
	\item Generate predicted values under  $\bar{a}_{i,\tau-2}^*$, and the observed $\bar{L}_{i,\tau-2}$ for all units uncensored at $\tau-2$ (i.e., $C_{i,\tau-2} = 0$).
	Let the predictions be denoted by $\tilde{Y}_{i,\tau-2}^*$.
	\item Repeat steps 3 and 4 for $\tilde{Y}_{i,j}^*$ where $j \in \{\tau-2, \tau-3, ..., 0\}$.
	\item Take the arithmetic mean of $\tilde{Y}_{i,0}^*$ across all $n$ observations.
\end{enumerate}
Step 6 provides the point estimate of $\mu_\tau(\bar{a}_{\tau-1}^{*})$, e.g., $\hat{\mu}_\tau(\bar{a}_{\tau-1}^{*}) = n^{-1} \sum_{i=1}^{n} \tilde{Y}_{i,0}^*$. Unlike standard g-computation, ICE g-computation only requires the analyst to specify $\tau$ regression models for $Y$. In other words, one does not need to specify models for other covariates included in $L_k$. For time-to-event data, the described ICE g-computation algorithm is slightly modified (see Appendix 1) \cite{wen_multiply_2022}.

The previous algorithm describes an unstratified version of ICE g-computation, meaning that the nuisance outcome models are fit to observations regardless of their observed treatment histories, $\bar{a}_{i,k-1}$. Therefore, the functional form for treatment history in each iterative regression model must be correctly specified. To avoid parametric assumptions for treatment history, the data used to estimate the outcome model parameters in steps 1 and 3 can instead be subset to those who followed the treatment plan $\bar{a}_{i,k-1}^*$ up to that time \cite{wen_parametric_2021}. We refer to this latter variation as stratified ICE g-computation. By estimating the models among only those following the plan, the stratified ICE g-computation estimator avoids any parametric constraints for the functional form of treatment history on the outcome. However, this comes at the cost of less data being available to fit the models.

For consistent variance estimation of $\hat{\mu}_\tau(\bar{a}_{\tau-1}^{*})$, the nonparametric bootstrap has been suggested \cite{schomaker_using_2019, wen_parametric_2021}. While the nonparametric bootstrap provides a consistent variance estimator, it can be computationally demanding, as it requires re-estimating the series of regression models using resampled data. Further, the ICE procedure does not provide estimates of $\mu_{\tau-1}(\bar{a}_{\tau-2}^{*})$ or the mean at other time-points as by-products of estimating $\mu_\tau(\bar{a}_{\tau-1}^{*})$, unlike standard g-computation. Instead, the whole process, and thus also the bootstrap, must be repeated for each time point of interest. Therefore, we consider an alternative to the bootstrap for estimating the variance of $\hat{\mu}_\tau(\bar{a}_{\tau-1}^{*})$.

\subsection{ICE g-computation as \edit{Estimating Equations}}

To avoid using the bootstrap to estimate the variance, we express the ICE g-computation estimator \edit{via estimating equations}. Let $\boldsymbol{\theta}$ be a $v$-dimensional vector, where $v$ is finite. The estimator $\hat{\boldsymbol{\theta}}$, \edit{often referred to as a M-estimator or Z-estimator} \cite{stefanski_calculus_2002, kosorok_z-estimators_2008}, is the solution to the estimating equation
\begin{equation*}
	\edit{\frac{1}{n}} \sum_{i=1}^{n} \psi(O_i; \hat{\boldsymbol{\theta}}) = \boldsymbol{0}
\end{equation*}
where $\psi(\cdot)$ is an estimating function \cite{stefanski_calculus_2002}. 
Given that an estimator solves a vector of unbiased estimating equations, it follows under suitable regularity that $\sqrt{n}(\hat{\boldsymbol{\theta}} - \boldsymbol{\theta}) \rightarrow^d N(0, \mathbb{V}(\boldsymbol{\theta}))$ \cite{boos_m-estimation_2013}, where 
\begin{equation*}
	\mathbb{V}(\boldsymbol{\theta}) = 
	\mathbb{B}(\boldsymbol{\theta})^{-1}
	\mathbb{F}(\boldsymbol{\theta})
	\left[\mathbb{B}(\boldsymbol{\theta})^{-1}\right]^T
\end{equation*}
is the sandwich variance, with the `bread' $\mathbb{B}(\boldsymbol{\theta}) = E\left[ - \partial \psi(O_i; \boldsymbol{\theta}) / \partial \boldsymbol{\theta}\right]$ and the `meat' $\mathbb{M}(\boldsymbol{\theta}) = E\left[\psi(O_i; \boldsymbol{\theta}) \psi(O_i; \boldsymbol{\theta})^T\right]$.
The asymptotic variance of $\hat{\boldsymbol{\theta}}$ can be consistently estimated by the empirical sandwich variance estimator, $\mathbb{V}_n(\hat{\boldsymbol{\theta}})$, consisting of the empirical counterparts for the bread matrix 
\begin{equation*}
	\mathbb{B}_n(\hat{\boldsymbol{\theta}}) = \frac{1}{n} \sum_{i=1}^{n} \left\{- \frac{\partial}{\partial \boldsymbol{\theta}} \psi(O_i; \hat{\boldsymbol{\theta}})\right\}
\end{equation*}
and meat matrix
\begin{equation*}
	\mathbb{M}_n(\hat{\boldsymbol{\theta}}) = \frac{1}{n} \sum_{i=1}^{n} \left\{\psi(O_i; \hat{\boldsymbol{\theta}}) \psi(O_i; \hat{\boldsymbol{\theta}})^T\right\}.
\end{equation*}
The covariance matrix $\mathbb{V}_{n}(\hat{\boldsymbol{\theta}})$ can then be used to compute the standard error for an element of $\hat{\boldsymbol{\theta}}$ and construct Wald-type confidence intervals (CI).

For unstratified ICE g-computation with a repeatedly measured binary outcome modeled using logistic regression, the \edit{estimating functions} are
\begin{equation}
	\label{equation3}
	\psi_U(O_i; \hat{\boldsymbol{\theta}}) = 
	\begin{bmatrix}
		I(C_{i, \tau} = 0) \left\{Y_{i,\tau} - \text{expit}(X_{i, \tau-1}^T \hat{\beta}_{\tau-1}) \right\}X_{i, \tau-1} \\
		I(C_{i, \tau-1} = 0) \left\{\tilde{Y}_{i,\tau-1}^* - \text{expit}(X_{i, \tau-2}^T \hat{\beta}_{\tau-2}) \right\}X_{i, \tau-2} \\
		\vdots \\
		I(C_{i, 1} = 0) \left\{\tilde{Y}_{i,1}^* - \text{expit}(X_{i, 0}^T \hat{\beta}_{0}) \right\}X_{i, 0} \\
		\tilde{Y}_{i,0}^* - \hat{\mu}_\tau(\bar{a}_{\tau-1}^{*})
	\end{bmatrix}
\end{equation}
where $\hat{\boldsymbol{\theta}} = (\hat{\beta}_{\tau-1}, \hat{\beta}_{\tau-2}, ... \hat{\beta}_{0}, \hat{\mu}_\tau(\bar{a}_{\tau-1}^{*}))$, $\text{expit}(\cdot)$ is the inverse logit function, $X_{i,k}$ is the $i$\textsuperscript{th} row of a design matrix composed of user-specified functions of $\bar{A}_{i,k}$ and $\bar{L}_{i,k}$, $\tilde{Y}_{i,k} = \text{expit}(X_{i,k-1}^{*^T} \hat{\beta}_{k-1})$, $X_{i,k}^*$ is the $i$\textsuperscript{th} row of the design matrix with $\bar{a}_{i,k}^*$ replacing $\bar{A}_{i,k}$, and $\hat{\mu}_\tau(\bar{a}_{\tau-1}^{*})$ is the estimator under the specified plan. The first estimating function is the score function of a logistic regression model for the observed outcome at time $\tau$. The second estimating function is the score of a fractional logistic regression model where the dependent variable is the predicted outcome at $\tau$ under the plan \cite{papke_econometric_1996}. The subsequent estimating functions are recursive fractional logistic regression models backwards through time until baseline. The final estimating function is for the mean of the outcome at $\tau$ under the plan. 

\edit{Following suitable regularity conditions} \cite[p.~327-329]{boos_m-estimation_2013}, one can show that $\hat{\boldsymbol{\theta}}$ is a consistent and asymptotically normal estimator of $\boldsymbol{\theta}$ by showing that $\hat{\boldsymbol{\theta}}$ is the solution to a vector of unbiased estimating equations. For the estimating equations corresponding to regression models, a generic element $x_{i, \tau-1}$ in the design matrix $X_{i, \tau-1}$ is considered without a loss of generality. 
For the first estimating equation in \ref{equation3}, note that it can be rewritten as an expectation conditional on $C_{\tau} = 0$, \edit{since the estimating function restricts contributions to observations with $C_{\tau} = 0$ via the leading indicator function}. Then it follows that  
\[E\left[\left\{Y_{i,\tau} - \text{expit}(X_{i, \tau-1}^T \beta_{\tau-1}) \right\} x_{i, \tau-1} \mid C_{\tau} = 0 \right] = 0 \]
given correct model specification (i.e., $E[Y_\tau \mid \bar{A}_{\tau-1}, \bar{L}_{\tau-1}, C_{\tau} = 0] = \text{expit}(X_{i, \tau-1}^T \beta_{\tau-1}) $), and maximum likelihood theory for logistic regression. Therefore, the first estimating equation is unbiased for $\beta_{\tau-1}$. For the second estimating equation in \ref{equation3}, it similarly follows that
\[E\left[ \left\{\tilde{Y}_{i,\tau-1}^* - \text{expit}(X_{i, \tau-2}^T \beta_{\tau-2}) \right\} x_{i, \tau-2} \mid C_{\tau - 1} = 0 \right] = 0\]
given the definition of $\tilde{Y}_{i,\tau-1}^*$ (i.e., $\tilde{Y}_{i,\tau-1}^* = E[Y_\tau \mid \bar{A}_{\tau-1} = \bar{a}^*_{\tau-1}, \bar{L}_{\tau-1}]$), correct model specification (i.e., $\text{expit}(X_{i, \tau-2}^T \beta_{\tau-2}) = E\left[E(Y_\tau \mid \bar{A}_{\tau-1} = \bar{a}_{\tau-1}^*, \bar{L}) \mid \bar{A}_{\tau-2}, \bar{L}_{\tau-2}\right]$), and from the quasi-maximum likelihood for fractional logistic regression \cite{papke_econometric_1996}. Therefore, the second estimating equation is unbiased for $\beta_{\tau-2}$. A similar argument for unbiasedness follows from correct model specification and the quasi-maximum likelihood for the subsequent estimating equations corresponding to $\beta_{\tau-3}, $$\beta_{\tau-4}, ..., \beta_{0}$. Finally, note that the last estimating equation is unbiased by the definitions of $\mu_\tau(\bar{a}^*_{\tau-1})$ and $\tilde{Y}^*_0$. Therefore, $\hat{\boldsymbol{\theta}}$ is a consistent and asymptotically normal estimator of $\boldsymbol{\theta}$.

The advantage of stacking the estimating equations is that all successive regression models for the ICE g-computation procedure are \textit{jointly} solved in \ref{equation3}. When \edit{finite-dimension} estimating equations are stacked together, standard M-estimation theory is applicable \cite{carroll_measurement_2006}. So, the variance of $\hat{\mu}_{\tau}(\bar{a}^*_{\tau-1})$ \edit{can be consistently estimated by} the empirical sandwich variance estimator \edit{when the estimating equations for all} nuisance model parameters \edit{are stacked together}. For intuition, notice that the bread matrix is defined as the partial derivatives of $\hat{\boldsymbol{\theta}}$. Similarly, the meat matrix is defined as the outer product of the stacked estimating equations. Therefore, the empirical sandwich variance estimator \edit{carries} the uncertainty \edit{across estimating functions included in the stack}.

For stratified ICE g-computation with a repeatedly measured binary outcome modeled using logistic regression, the estimating \edit{functions} are instead 
\begin{equation}
	\label{equation4}
	\psi_S(O_i; \check{\boldsymbol{\theta}}) = 
	\begin{bmatrix}
		I(C_{i, \tau} = 0, \bar{A}_{i,\tau-1} = \bar{a}_{i,\tau-1}^*) \left\{Y_{i,\tau} - \text{expit}(X_{i, \tau-1}^T \check{\beta}_{\tau-1}) \right\}X_{i, \tau-1} \\
		I(C_{i, \tau-1} = 0, \bar{A}_{i,\tau-2} = \bar{a}_{i,\tau-2}^*) \left\{\tilde{Y}_{i,\tau-1}^* - \text{expit}(X_{i, \tau-2}^T \check{\beta}_{\tau-2}) \right\}X_{i, \tau-2} \\
		\vdots \\
		I(C_{i, 1} = 0, \bar{A}_{i,0} = \bar{a}_{i,0}^*) \left\{\tilde{Y}_{i,1}^* - \text{expit}(X_{i, 0}^T \check{\beta}_{0}) \right\}X_{i, 0} \\
		\tilde{Y}_{i,0}^* - \check{\mu}_\tau(\bar{a}_{\tau-1}^{*})
	\end{bmatrix}
\end{equation}
where $\check{\boldsymbol{\theta}} = (\check{\beta}_{\tau-1}, \check{\beta}_{\tau-2}, ... \check{\beta}_{0}, \check{\mu}_\tau(\bar{a}_{\tau-1}^{*}))$. These estimating equations are nearly identical to Equation \ref{equation3}, except for the addition of being restricted to those who followed the treatment plan up to the corresponding time. Notice that a similar proof of unbiasedness for the stacked estimating equations follows here, with stratified ICE g-computation only relaxing the parametric modeling assumptions being made.

The score functions of the logistic models in Equations \ref{equation3} and \ref{equation4} can also be replaced with the score functions of other generalized linear models. As indicated above, inference for $\hat{\mu}_\tau(\bar{a}_{\tau-1}^{*})$ and $\check{\mu}_\tau(\bar{a}_{\tau-1}^{*})$ can then be made using the empirical sandwich variance estimator. Importantly, software is available which automates computation of the point and variance estimates for a given set of stacked estimating equations \cite{zivich_delicatessen_2022, saul_calculus_2020}.

\section{Simulation study}

To explore the finite-sample performance of the empirical sandwich variance estimator against theoretical expectations, a simulation study was conducted. The parameters of interest in the simulations were $\mathbb{E}[Y_{i,3}(1, 1, 1)]$ and $\mathbb{E}[Y_{i,3}(0, 0, 0)]$, which correspond to the mean had everyone been treated at all three time points and the mean had everyone not been treated at all three time points, respectively. 
Here, $L_{i,k}$ consisted of a binary variable generated via
\[L_{i,0} \sim \text{Bernoulli}(0.5)\]
\[L_{i,1}(\bar{a}_0) \sim \text{Bernoulli}\left\{\text{expit}(-1 - a_0 + L_{i,0})\right\}\]
\[L_{i,2}(\bar{a}_1) \sim \text{Bernoulli}\left\{\text{expit}(-1 - 0.2 a_0 - a_1 + 0.5 L_{i,0} + L_{i,1}(a_0))\right\}\]
so that $L_{i,k}$ depended on prior values of $L$ and previous treatments. Notice that $L_1$ and $L_2$ are defined as potential outcomes that depend on $\bar{a}$. Potential outcomes for $Y$ were generated by the following models
\[Y_{i,1}(\bar{a}_0) \sim \text{Bernoulli}\left\{\text{expit}(-1.5 + 0.5 a_0 - 2 L_{i,0})\right\}\]
\[Y_{i,2}(\bar{a}_1) \sim \text{Bernoulli}\left\{\text{expit}(-1.5 + 0.1 a_0 + 1.2 a_1 - 0.5 L_{i,0} - 2 L_{i,1}(a_0))\right\}\]
\[Y_{i,3}(\bar{a}_2) \sim \text{Bernoulli}\left\{\text{expit}(-1.5 + 0.1 a_1 + 1.2 a_2 - 0.5 L_{i,1}(\bar{a}_0) - 2 L_{i,2}(\bar{a}_1))\right\}.\]
The observed values of $L$ and $Y$ were generated from the potential outcomes via causal consistency, defined for a random variable, $Z$, as 
\[Z_{i,t+1} = \sum_{\bar{a}_{t}} Z_{i,t+1}(\bar{a}_{t}) I(A_{i, t} = \bar{a}_{t})\]
and the observed treatments were generated as follows
\[A_{i,0} \sim \text{Bernoulli}\left\{\text{expit}(1 - 2 L_{i,0})\right\}\]
\[A_{i,1} \sim \text{Bernoulli}\left\{\text{expit}(-1 - 0.2 L_{i,0} - L_{i,1} + 1.75 A_{i,0})\right\}\]
\[A_{i,2} \sim \text{Bernoulli}\left\{\text{expit}(-1 - 0.2 L_{i,1} - L_{i,2} + 1.75 A_{i,1})\right\}\]
so that $L$ was a time-varying confounder. Further, monotonic informative loss to follow-up was induced by
\[C_{i,1} \sim \text{Bernoulli}\left\{\text{expit}(-2.5 - 0.5 A_{i,0})\right\}\]
\[C_{i,2} \sim 
\begin{cases}
	1 & \text{if } C_{i,1} = 1 \\
	\text{Bernoulli}\left\{\text{expit}(-2.5 - 0.5 A_{i,1})\right\} & \text{if } C_{i,1} = 0 \\
\end{cases}\]
\[C_{i,3} \sim 
\begin{cases}
	1 & \text{if } C_{i,2} = 1 \\
	\text{Bernoulli}\left\{\text{expit}(-2.5 - 0.5 A_{i,2})\right\} & \text{if } C_{i,2} = 0 \\
\end{cases}\]
where $\left\{L_{i,k+1}, A_{i,k+1}, Y_{i,k}\right\}$ were set to missing if $C_{i,k} = 1$.

For estimation of the parameters of interest, unstratified and stratified ICE g-computation \edit{estimating equations} under correct model specification were assessed. \edit{Estimators} were evaluated at five difference sample sizes, $n \in \{250, 500, 1000, 2000, 5000\}$, with 5000 iterations each. The following metrics were evaluated: bias, empirical standard error (ESE), average standard error (ASE), standard error ratio (SER), and 95\% CI coverage \cite{morris_using_2019}. Bias was defined as the mean of the estimated mean minus the true mean under the plan, with the true mean determined by simulating 10 million observations under the plan. ESE was estimated by the standard deviation of the point estimates of the simulation. ASE was estimated by the mean of the estimated standard errors. SER was defined as the ASE divided by the ESE. CI coverage was estimated by the proportion of 95\% CI that contained the true mean under the plan. Whether the \edit{root-finding} procedure failed within 10000 iterations was also tracked. Failures to converge were ignored for computation of the other metrics.

Simulations were conducted using Python 3.9.4 (Python Software Foundation, Beaverton, OR, USA) with the following packages: \texttt{NumPy} \cite{harris_array_2020}, \texttt{SciPy} \cite{virtanen_scipy_2020}, \texttt{delicatessen} \cite{zivich_delicatessen_2022}, and \texttt{pandas} \cite{mckinney_data_2010}. Code to replicate the simulation results is provided at \url{github.com/pzivich/publications-code}.

\subsection{Results}

Simulation results are presented in Tables 2-3. As seen across sample sizes, both unstratified and stratified ICE g-computation \edit{estimators} were approximately unbiased under correct model specification. In general, unstratified ICE g-computation has a smaller ESE, and thus was more precise, relative to stratified ICE g-computation. However, the difference between the ESE of the stratified and unstratified approaches diminished as sample sizes increased. For small sample sizes, like $n=250$, stratified ICE g-computation occasionally failed to converge (up to 6\%). Both the differences in precision and failure to converge are likely due to data becoming sparse (i.e., few observations followed the plan of interest and thus the nuisance models either could not be estimated or had large variances). These issues were more pronounced for the always-treat parameter as treatment was less prevalent at all time points in the data generating mechanism.

Performance of the empirical sandwich variance estimator aligned with theoretical expectations. Across the varying sample sizes, the SER was near 1 and CI coverage was near 0.95 for both estimators. However, the sandwich variance estimator under performed for the smallest sample size ($n=250$), most likely attributable to the data sparsity previously noted.

\begin{table}[]
	\caption{ICE g-formula always-treat simulation results}
	\centering
	\begin{tabular}{llcccccc}
		\hline
		&                    & Bias   & ESE   & ASE   & SER  & Coverage & Failed\textsuperscript{a} \\ \hline
		\multicolumn{2}{l}{$n=250$}  &        &       &       &      &          &        \\
		& Unstratified       & -0.003 & 0.061 & 0.061 & 1.01 & 0.94     & 0      \\
		& Stratified         & 0.005  & 0.081 & 0.074 & 0.91 & 0.91     & 288    \\
		\multicolumn{2}{l}{$n=500$}  &        &       &       &      &          &        \\
		& Unstratified       & -0.002 & 0.043 & 0.043 & 1.00 & 0.94     & 0      \\
		& Stratified         & 0.003  & 0.056 & 0.054 & 0.97 & 0.94     & 55     \\
		\multicolumn{2}{l}{$n=1000$} &        &       &       &      &          &        \\
		& Unstratified       & -0.004 & 0.031 & 0.031 & 0.99 & 0.95     & 0      \\
		& Stratified         & 0.001  & 0.039 & 0.038 & 0.99 & 0.95     & 1      \\
		\multicolumn{2}{l}{$n=2000$} &        &       &       &      &          &        \\
		& Unstratified       & -0.004 & 0.021 & 0.022 & 1.01 & 0.95     & 0      \\
		& Stratified         & 0.001  & 0.027 & 0.027 & 1.00 & 0.95     & 0      \\
		\multicolumn{2}{l}{$n=5000$} &        &       &       &      &          &        \\
		& Unstratified       & -0.004 & 0.014 & 0.014 & 0.99 & 0.94     & 0      \\
		& Stratified         & 0.001  & 0.017 & 0.017 & 0.99 & 0.95     & 0      \\ \hline
	\end{tabular}
	\floatfoot{
		ESE: empirical standard error, ASE: average standard error, SER: standard error ratio (ASE/ESE), Coverage: 95\% confidence interval (CI) coverage. The parameter of interest in the simulation was $\mathbb{E}[Y_{3}(1,1,1)]$. \\
		Bias was defined as the mean of the estimated mean minus the true mean had everyone been treated at all time points. ESE: was defied as the standard deviation of the simulation estimates. ASE was the mean of the estimated standard errors across all simulations. SER was the ASE divided by the ESE. 95\% CI coverage was defined as the proportion of 95\% CIs containing the true mean. Failed indicates whether the root-finding procedure failed to converge in 10,000 iterations. Failed iterations were ignored for calculation of other metrics. Results are for 5000 iterations. \\
		\textsuperscript{a} Failed convergences were ignored when calculating the evaluation metrics.
	}
\end{table}

\begin{table}[]
	\caption{ICE g-formula never-treat simulation results}
	\centering
	\begin{tabular}{llcccccc}
		\hline
		&                    & Bias   & ESE   & ASE   & SER  & Coverage & Failed \\ \hline
		\multicolumn{2}{l}{$n=250$}  &        &       &       &      &          &        \\
		& Unstratified       & -0.002 & 0.034 & 0.033 & 0.97 & 0.93     & 0      \\
		& Stratified         & 0.000  & 0.047 & 0.046 & 0.98 & 0.92     & 13     \\
		\multicolumn{2}{l}{$n=500$}  &        &       &       &      &          &        \\
		& Unstratified       & -0.001 & 0.023 & 0.023 & 1.00 & 0.94     & 0      \\
		& Stratified         & 0.001  & 0.033 & 0.033 & 0.99 & 0.93     & 0      \\
		\multicolumn{2}{l}{$n=1000$} &        &       &       &      &          &        \\
		& Unstratified       & -0.001 & 0.017 & 0.017 & 1.00 & 0.94     & 0      \\
		& Stratified         & 0.000  & 0.023 & 0.023 & 1.00 & 0.95     & 0      \\
		\multicolumn{2}{l}{$n=2000$} &        &       &       &      &          &        \\
		& Unstratified       & -0.001 & 0.012 & 0.012 & 1.00 & 0.94     & 0      \\
		& Stratified         & 0.000  & 0.017 & 0.016 & 0.99 & 0.94     & 0      \\
		\multicolumn{2}{l}{$n=5000$} &        &       &       &      &          &        \\
		& Unstratified       & -0.001 & 0.007 & 0.007 & 1.01 & 0.95     & 0      \\
		& Stratified         & 0.000  & 0.010 & 0.010 & 1.00 & 0.95     & 0      \\ \hline
	\end{tabular}
	\floatfoot{
		ESE: empirical standard error, ASE: average standard error, SER: standard error ratio (ASE/ESE), Coverage: 95\% confidence interval (CI) coverage. The parameter of interest in the simulation was $\mathbb{E}[Y_{3}(0,0,0)]$. \\
		Bias was defined as the mean of the estimated mean minus the true mean had no one been treated at all time points. ESE: was defied as the standard deviation of the simulation estimates. ASE was the mean of the estimated standard errors across all simulations. SER was the ASE divided by the ESE. 95\% CI coverage was defined as the proportion of 95\% CIs containing the true mean. Failed indicates whether the root-finding procedure failed to converge in 10,000 iterations. Failed iterations were ignored for calculation of other metrics. Results are for 5000 iterations. \\
		\textsuperscript{a} Failed convergences were ignored when calculating the evaluation metrics.
	}
\end{table}

\section{Example}

As an illustration of ICE g-computation, we estimate the prevalence of hypertension had all cigarette smoking been prevented among adolescents enrolled in school for the 1994-1995 academic year. While preventing cigarette smoking is not a `treatment', the framework of the previous sections also applies to exposures, interventions, and actions more generally. Data came from the Add Health public-use in-home questionnaire ($n=6504$), which removes some variables and constitutes a subsample of the full Add Health data to limit potential for deductive disclosures. Add Health is a school-based, nationally representative study of adolescents in grades 7-12 that began in 1994-1995. After wave I (1994-1995), additional waves have been conducted in 1996 (wave II), 2001-2002 (wave III), and 2008-2009 (wave IV). Details on the design of Add Health are available in Harris et al. (2019) \cite{harris_cohort_2019}. To maintain similar lengths of time between follow-up visits (e.g., approximately seven years), only data from waves I, III, and IV were used. The following exclusion criteria were applied: no self-reported heart problem resulting in difficulty using your hands, arms, legs, or feet at wave I (underlying health conditions potentially related to elevated blood pressure that were not commonly observed); between 13-18 years old at wave I (to prevent data sparsity by age); in a high school with grade levels at wave I (for definition of education at wave I); and best described race was not `other' at wave I (`other' was not an option for wave III).

The parameters of interest for our analysis was the prevalence of hypertension at wave IV had none of the defined population been current cigarette smokers at waves I and III, $\mu_2(0,0) = \mathbb{E}[Y_{i,2}(0, 0)]$. Additionally, we assessed the causal effect of this smoking ban relative to the natural course (i.e., observed patterns of smoking) \cite{rudolph_role_2021}, $\delta_{2} = \mu_2(0, 0) - \mu_2$ where $\mu_{2} = \mathbb{E}[Y_{i,2}]$. At wave IV, participants with type I (systolic between 140-159 or diastolic between 90-99) or type II (systolic 160+ or diastolic 100+) hypertension as measured by the interviewer were classified as having hypertension. For each visit, current cigarette smoking was defined as those who reported smoking cigarettes at least one day in the previous 30 days from the interview data. Those who reported zero days or never smoking were classified as not current cigarette smokers.

For identification, we assumed sequential exchangeability of treatment and censoring given the following time-varying and time-fixed covariates. Time-varying confounders included gender, race, ethnicity, height, weight, exercise, self-rated health, alcohol use, prior hypertension, and health insurance coverage. While gender, race, and ethnicity are not traditionally considered to be time-varying covariates, there has been increasing recognition that these characteristics can vary over time \cite{tabb_changes_2016, agadjanian_how_2022, collin_prevalence_2016}. Time-fixed confounders include age and ever trying a cigarette at wave I. Further details on variable definitions are provided in Appendix 2. To construct the analytic data set, those with any missing covariates at wave I were excluded. To ensure missing data was monotonic across time, a participant's covariates were all set to missing if any of their covariates at previous waves were missing.

For estimation of the mean under a smoking ban, the unstratified ICE g-computation estimator, $\hat{\mu}_2(0,0)$, was used. Rather than calculating the mean under the natural course directly, we instead used unstratified ICE g-computation with the observed values of $\bar{A}_{i,2}$. This approach has the advantage of accounting for informative loss to follow-up by $(\bar{A}_{i,k},\bar{L}_{i,k})$, unlike simply taking the mean of the observed $Y_2$. The ICE g-computation estimators for each plan were stacked together. Finally, an estimating function for the difference between the proportion under the smoking ban and natural course was stacked as well. The full set of stacking estimating equations for $\hat{\delta}_{2}$ were
\begin{equation*}
	\sum_{i=1}^{n} 
	\begin{bmatrix}
		I(C_{i, 2} = 0) \left\{Y_{i,2} - \text{expit}(X_{i, 1}^T \hat{\beta}_{1}) \right\}X_{i,1} \\
		I(C_{i, 1} = 0) \left\{\tilde{Y}_{i,1}^* - \text{expit}(X_{i, 0}^T \hat{\beta}_{0}) \right\}X_{i, 0} \\
		\tilde{Y}_{i,0}^* - \hat{\mu}_2(0,0) \\

		I(C_{i, 1} = 0) \left\{\tilde{Y}_{i,1}^{*'} - \text{expit}(X_{i, 0}^T \hat{\beta}_{0}) \right\}X_{i, 0} \\
		\tilde{Y}_{i,0}^{*'} - \hat{\mu}_{2} \\

		(\hat{\mu}_2(0,0) - \hat{\mu}_2) - \hat{\delta}_{2}
	\end{bmatrix}
	= \boldsymbol{0}
\end{equation*}
where $\tilde{Y}_{i,k}^*$ is the predicted hypertension probability under the cigarette smoking ban and $\tilde{Y}_{i,k}^{*'}$ is the predicted hypertension probability under the observed value of cigarette smoking ban at time $k$.

The following specifications were used for outcome models. Height and weight were rescaled to be standard normal, and then modeled using restricted cubic splines with knots located at the 5\textsuperscript{th}, 33\textsuperscript{rd}, 67\textsuperscript{th}, and 95\textsuperscript{th} percentiles. Age, exercise, alcohol, self-rated health, and education were modeled as disjoint indicator terms. The outcome model for hypertension at wave IV included baseline and time-varying variables from wave III only, expect for current cigarette smoking which included both wave I and wave III smoking status. No interaction terms were included in models besides an interaction term between wave I and wave III smoking status for hypertension at wave IV. The second model of each ICE g-computation only included variables from wave I.

Analyses were conducted with Python 3.9.4 and replicated in R 4.2.0 (Vienna, Austria) with the \texttt{geex}, \texttt{numDeriv}, and \texttt{rootSolve} packages \cite{saul_calculus_2020,varadhan_numderiv_2019, soetaert_rootsolve_2021}. Data are freely available from the University of North Carolina at Chapel Hill Dataverse hosted by the Odum Institute \cite{harris_national_2020, harris_national_2020-1, harris_national_2020-2}. Code to preprocess the data set and replicate the analysis is provided at \url{gihtub.com/pzivich/publications-code}. To compare runtimes, several implementations of ICE g-computation were considered. As a reference, ICE g-computation was implemented using successive generalized linear models, with the variance estimated using a nonparametric bootstrap with 500 resamples. Point estimates for nuisance models were computed using iteratively reweighted least squares with \texttt{statsmodels} \cite{seabold_statsmodels_2010}. Bootstrap iterations were run both in sequence and up to seven in parallel. Two implementations of \edit{the estimating equations} were compared. First, the solution $\hat{\boldsymbol{\theta}}$ was found via root-finding with the Levenberg-Marquardt with generic starting values (i.e., $0.18$ for the causal means and $0.0$ for all other parameters). As this approach involves the simultaneous estimation of many parameters, the root-finding procedure can take may iterations to converge. So, an implementation that first solves for the nuisance parameters using iteratively reweighted least squares and then uses Levenberg-Marquardt to solve for the parameters of interest was also considered. While the reported runtime still includes solving for the nuisance parameters subsets, the overall runtime is expected to be lower. Runtime results were reported for Python only.

\subsection{Results}

After application of the exclusion criteria, 5657 (87\%) observations remained in the analytic data set. Descriptive statistics for the analytic data set are provided in Tables 4-5. Between waves I and III, 1694 (30\%) observations were censored. Between waves III and IV, an additional 594 (15\%) of observations were censored. Had all current smoking been prevented at waves I and III, the estimated prevalence of hypertension at wave IV would have been 0.175 (95\% CI: 0.157, 0.193), which is 1.15 percentage points lower (95\% CI: -0.025, 0.002) than the natural course. The sandwich variance estimator provided a similar CI to the bootstrap but was substantially faster (Table 6). This result remained true even when up to seven bootstrap iterations were run in parallel. The reduction in runtime was the greatest when the nuisance parameters were solved prior to root-finding, indicating most of the runtime with generic starting values was spent finding the roots of the estimating equations. Altogether, this suggests that implementations with a bit more involved coding can greatly reduce runtimes, but even out-of-the-box implementations can still reduce runtimes.

\begin{table}[]
	\caption{Descriptive Statistics for Add Health Wave I}
	\begin{tabular}{llc} \hline
		&                                       & Wave I ($n=5657$)  \\ \hline
		\multicolumn{2}{l}{Current cigarette smoker\textsuperscript{a}}          & 1462 (26\%)        \\
		\multicolumn{2}{l}{Age}                               & 16 {[}15, 17{]}    \\
		\multicolumn{2}{l}{Ever tried smoking a cigarette}    & 3141 (56\%)        \\
		\multicolumn{2}{l}{Female}                            & 2779 (49\%)        \\
		\multicolumn{2}{l}{Race}                              &                    \\
		& White                                 & 3911 (69\%)        \\
		& Black                                 & 1434 (25\%)        \\
		& Native American                       & 90 (2\%)           \\
		& Asian or Pacific Islander             & 222 (4\%)          \\
		\multicolumn{2}{l}{Hispanic}                          & 358 (6\%)          \\
		\multicolumn{2}{l}{Current grade level}               &                    \\
		& 7\textsuperscript{th}                 & 851 (15\%)         \\
		& 8\textsuperscript{th}                 & 882 (16\%)         \\
		& 9\textsuperscript{th}                 & 997 (18\%)         \\
		& 10\textsuperscript{th}                & 1027 (18\%)        \\
		& 11\textsuperscript{th}                & 1019 (18\%)        \\
		& 12\textsuperscript{th}                & 881 (16\%)         \\
		\multicolumn{2}{l}{Height (inches)}                   & 66 {[}63, 69{]}    \\
		\multicolumn{2}{l}{Weight (pounds)}                   & 135 {[}118, 160{]} \\
		\multicolumn{2}{l}{Alcohol use in prior 12 months}    &                    \\
		& Never                                 & 3048 (54\%)        \\
		& 1-2 days total                        & 969 (17\%)         \\
		& 1-3 times a month                     & 1106 (20\%)        \\
		& 1-2 times a week                      & 343 (6\%)          \\
		& 3 or more times a week                & 191 (3\%)          \\
		\multicolumn{2}{l}{Exercise over previous seven days} &                    \\
		& None                                  & 917 (16\%)         \\
		& 1-2 times                             & 1797 (32\%)        \\
		& 3-4 times                             & 1400 (25\%)        \\
		& 5 or more times                       & 1543 (27\%)        \\
		\multicolumn{2}{l}{Self-rated health}                 &                    \\
		& Excellent                             & 1637 (29\%)        \\
		& Very good                             & 2299 (41\%)        \\
		& Good                                  & 1369 (24\%)        \\
		& Fair                                  & 332 (6\%)          \\
		& Poor                                  & 20 (0\%)           \\ \hline
	\end{tabular}
	\floatfoot{
		\textsuperscript{a} Defined as those who reported smoking cigarettes for at least one day in the previous 30 days from the interview date. Those who reported zero days or never smoking were classified as not current cigarette smokers. 
	}
\end{table}

\begin{table}[]
	\caption{Descriptive Statistics for Add Health Wave III}
	\begin{tabular}{llc}
		\hline
		&                                         & Wave III ($n=3963$) \\ \hline
		\multicolumn{2}{l}{Current cigarette smoker}              & 1322 (33\%)         \\
		\multicolumn{2}{l}{Female}                                & 1882 (47\%)         \\
		\multicolumn{2}{l}{Race}                                  &                     \\
		& White                                   & 2764 (70\%)         \\
		& Black                                   & 970 (24\%)          \\
		& Native American                         & 69 (2\%)            \\
		& Asian or Pacific Islander               & 160 (4\%)           \\
		\multicolumn{2}{l}{Hispanic}                              & 232 (6\%)           \\
		\multicolumn{2}{l}{Highest grade completed}               &                     \\
		& Less than high school                   & 451 (11\%)          \\
		& High school                             & 1220 (31\%)         \\
		& At least some college                   & 2207 (56\%)         \\
		& Pursuit of graduate degree              & 85 (2\%)            \\
		\multicolumn{2}{l}{Height (inches)}                       & 67 {[}64, 70{]}     \\
		\multicolumn{2}{l}{Weight (pounds)}                       & 163 {[}138, 194{]}  \\
		\multicolumn{2}{l}{Alcohol use in prior 12 months}        &                     \\
		& Never                                   & 1048 (26\%)         \\
		& 1-2 days total                          & 444 (11\%)          \\
		& 1-3 times a month                       & 1293 (33\%)         \\
		& 1-2 times a week                        & 801 (20\%)          \\
		& 3 or more times a week                  & 377 (10\%)          \\
		\multicolumn{2}{l}{Exercise over previous seven days}     &                     \\
		& None                                    & 782 (20\%)          \\
		& 1-2 times                               & 664 (17\%)          \\
		& 3-4 times                               & 597 (15\%)          \\
		& 5 or more times                         & 1920 (48\%)         \\
		\multicolumn{2}{l}{Self-rated health}                     &                     \\
		& Excellent                               & 1317 (33\%)         \\
		& Very good                               & 1662 (42\%)         \\
		& Good                                    & 812 (20\%)          \\
		& Fair                                    & 156 (4\%)           \\
		& Poor                                    & 16 (0\%)            \\
		\multicolumn{2}{l}{No or unknown health insurance status} & 890 (22\%)          \\
		\multicolumn{2}{l}{Ever diagnosed with HPB or HTN}        & 231 (6\%)           \\ \hline
	\end{tabular}
	\floatfoot{
		HBP: high blood pressure, HTN: hypertension. \\
		\textsuperscript{a} Defined as those who reported smoking cigarettes for at least one day in the previous 30 days from the interview date. Those who reported zero days or never smoking were classified as not current cigarette smokers. 
	}
\end{table}

We note that this analysis should only be viewed as an illustration of ICE g-computation, as the identification assumptions are unlikely to be reasonably met in this example. Specifically, other uses of smoking tobacco were ignored. Socio-economic status and diet were not included in the adjustment set, a likely violation of treatment exchangeability. Measurement error of self-reported covariates is probable, particularly for self-reported cigarette use. The analysis also ignored the Add Health sampling weights, so inference to the stated Add Health target population is not appropriate. To incorporate Add Health sampling weights, we would replace each of the estimating equations for the logistic models and arithmetic means with their sample weighted counterparts. Finally, follow-up visits were every seven years. As described elsewhere \cite{ferreira_guerra_impact_2020}, how data is discretized or coarsened over time can result in a loss of information regarding time-varying covariates, which can lead to bias. As such, the follow-up design of Add Health may produce bias in the estimate for the intervention and outcome considered in this analysis.

\begin{table}[]
	\caption{Results for the illustrative example of cigarette smoking on prevalent hypertension using data from Add Health}
	\centering
	\begin{tabular}{llcc}
		\hline
		\multicolumn{2}{l}{Prevent cigarette smoking}       & Risk (95\% CI)         & Runtime\textsuperscript{a} \\
		& Sandwich                           & 0.175 (0.157, 0.193)   & 5.7     \\
		& Sandwich -- pre-solving\textsuperscript{b}                    & 0.175 (0.157, 0.193)   & 0.8     \\
		& Bootstrap in sequence\textsuperscript{c}              & 0.175 (0.156, 0.194)   & 40.3    \\
		& Bootstrap in parallel\textsuperscript{d}              & 0.175 (0.156, 0.194)   & 21.2    \\
		&                                    &                        &         \\
		\multicolumn{2}{l}{Effect of cigarette smoking ban} & Ban effect (95\% CI)   & Runtime\textsuperscript{a} \\
		& Sandwich                           & -0.011 (-0.025, 0.002) & 12.9    \\
		& Sandwich -- pre-solving\textsuperscript{b}                    & -0.011 (-0.025, 0.002) & 2.7     \\
		& Bootstrap in sequence\textsuperscript{c}              & -0.011 (-0.025, 0.002) & 55.7    \\
		& Bootstrap in parallel\textsuperscript{d}              & -0.011 (-0.025, 0.002) & 26.3    \\ \hline
	\end{tabular}
	\floatfoot{
		95\% CI: 95\% confidence interval. \\
		\textsuperscript{a} Reported in seconds. Run-times were determined using a laptop running Windows 10 Pro on the ‘best performance’ power setting with an 11th generation Intel® 4-core 2.60 GHz processor and 16GB of RAM.\\
		\textsuperscript{b} Results are based on solving the nuisance parameter estimates using iteratively reweighted least squares and root-finding for only the parameters of interest. This process was expected to reduce the overall runtime related to the other reported sandwich results, which involved simultaneous estimation of all parameters via root-finding. \\
		\textsuperscript{c} Nonparametric bootstrap results based on the standard deviation of 500 resamples. Run-times are for bootstraps run in sequence (i.e., one bootstrap iteration must be completed before the next is ran). \\
		\textsuperscript{d} Nonparametric bootstrap results based on the standard deviation of 500 resamples. Run-times are for seven bootstraps run in parallel. While possible to run more bootstrap resamples in parallel, further reductions in run-time depend on the availability of computational resources (e.g., number of cores, RAM).
	}
\end{table}

\section{Conclusions}

Here, we expressed the ICE g-computation estimator \edit{as a set of estimating equations} to reduce the computational burden of variance estimation with the nonparametric bootstrap. Performance of the empirical sandwich variance estimator in the simulation study aligned with expectations and provided notable reductions in runtimes in the applied example. As indicated by our simulations, stratified ICE g-computation may fail to converge or have poor performance with small sample sizes. In these cases, unstratified ICE g-computation may be preferred, under the additional assumption that the parametric constraints used in the models are deemed to be close approximations.

This paper focused on deterministic plans that did not depend on the natural course. However, ICE g-computation has been extended for plans that depend on the natural course or stochastic plans (i.e., plans where treatment is assigned probabilistically) \cite{wen_intervention_2023}. Generalizations of the proposed \edit{estimating equations} could also be developed for these extensions of ICE g-computation. Use of \edit{estimating equations} is also not limited to g-computation. Inverse probability weighting estimators for longitudinal data can be expressed as stacked estimating equations \cite{wen_multiply_2022}, thereby avoiding the conservative estimation of the variance via the `robust' variance estimator in some settings \cite{robins_marginal_2000, reifeis_variance_2022}. Multiply-robust estimators can also be expressed as estimating equations \cite{kreif_estimating_2017, schomaker_using_2019, wen_multiply_2022, bang_doubly_2005, petersen_targeted_2014}. For some versions of these multiply-robust estimators, closed-form variance estimators based on the influence curve are available \cite{bang_doubly_2005, petersen_targeted_2014}. \edit{However, the empirical sandwich variance estimator may offer multiply-robust inference, unlike the influence curve variance estimators \cite{shook_double_2024}.}
\edit{Therefore}, the empirical sandwich variance estimator remains an appealing option. 

\section*{Acknowledgments}

Conflicts of Interest: None to declare. Financial Support: This work was supported in part by T32AI007001 (PNZ), K01AI177102 (PNZ), R01AI157758 (SRC, JKE, BES). Data and Code: Data used for the illustrative example are publicly available from the University of North Carolina at Chapel Hill Dataverse hosted by the Odum Institute. Code to replicate the illustrative example and the simulation experiment is available at \\
\url{https://github.com/pzivich/publications-code}.

\small
\bibliography{biblio}{}
\bibliographystyle{ieeetr}

\newpage 

\section*{Appendix}

\subsection*{Appendix 1: Iterated conditational expectation for time-to-event data}

For time-to-event data, the previously described algorithm for ICE g-computation is modified. Specifically, the predicted outcome is replaced with the observed outcome if the event occurred at time $k$ for unit $i$ in steps 2 and 4, and models are fit using only the units who survived up to the start of that interval. The following algorithm can be used to implement the ICE g-computation estimator for time-to-event data:
\begin{enumerate}
	\item Fit a regression model for $Y_{i,\tau}$ conditional on $\bar{A}_{i,\tau-1}$ and $\bar{L}_{i,\tau-1}$ for all observations where $C_{i,\tau} = 0$ and $Y_{i,\tau-1} = 0$.
	\item Generate predicted values of the outcome for the plan of interest, $\bar{a}_{i,\tau-1}^*$, and the observed $\bar{L}_{i,\tau-1}$ for all units uncensored at $\tau-1$ (i.e., $C_{i,\tau-1} = 0$). Let the predictions be denoted $\tilde{Y}_{i,\tau-1}^*$. If $Y_{i,\tau-1} = 1$ then set $\tilde{Y}_{i,\tau-1}^* = 1$.
	\item Fit a regression model for $\tilde{Y}_{i,\tau}^*$ conditional on $\bar{A}_{i,\tau-2}$ and $\bar{L}_{i,\tau-2}$ with all observations where $C_{i,\tau-1} = 0$ and $Y_{i,\tau-2} = 0$.
	\item Generate predicted values of the outcome under $\bar{a}_{i,\tau-2}^*$, and the observed $\bar{L}_{i,\tau-2}$ for all units uncensored at $\tau-2$ (i.e., $C_{i,\tau-2} = 0$). Let the predictions be denoted $\tilde{Y}_{i,\tau-2}^*$. If $Y_{i,\tau-2} = 1$ then set $\tilde{Y}_{i,\tau-2}^* = 1$.
	\item Repeat steps 3 and 4 for $\tilde{Y}_{i,j}^*$ where $j \in \{\tau-1, \tau-2, ..., 1\}$.
	\item Take the arithmetic mean of $\tilde{Y}_{i,1}^*$ across all $n$ observations.
\end{enumerate}
For survival endpoints, the corresponding estimating equations for the unstratified ICE g-computation estimator are:
\begin{equation*}
	\sum_{i=1}^{n} \psi_U(O_i; \hat{\boldsymbol{\theta}}) = 
	\sum_{i=1}^{n} 
	\begin{bmatrix}
		I(C_{i, \tau} = 0, Y_{i, \tau-1} = 0) \left\{Y_{i,\tau} - \text{expit}(X_{i, \tau-1}^T \hat{\beta}_{\tau-1}) \right\}X_{i, \tau-1} \\
		I(C_{i, \tau-1} = 0, Y_{i, \tau-2} = 0) \left\{\tilde{Y}_{i,\tau-1}^* - \text{expit}(X_{i, \tau-2}^T \hat{\beta}_{\tau-2}) \right\}X_{i, \tau-2} \\
		\vdots \\
		I(C_{i, 1} = 0) \left\{\tilde{Y}_{i,1}^* - \text{expit}(X_{i, 0}^T \hat{\beta}_{0}) \right\}X_{i, 0} \\
		\tilde{Y}_{i,0}^* - \mu_\tau(\bar{a}_{\tau-1}^{*})
	\end{bmatrix}
	= \boldsymbol{0}
\end{equation*}
where $\tilde{Y}_{i,k}^*$ is defined as 
\begin{equation*}
	\tilde{Y}_{i,k}^* = 
	\begin{cases}
		\text{expit}(X_{i,k}^{*^T} \hat{\beta}_{k}) & \text{if } Y_{i,k} \ne 1\\
		1              & \text{if } Y_{i,k} = 1
	\end{cases}
\end{equation*}
The estimating equation for stratified ICE g-computation can be implemented by replacing $I(C_{i, k} = 0, Y_{i, k-1} = 0)$ with $I(C_{i, k} = 0, Y_{i, k-1} = 0, \bar{A}_{i,k-1} = \bar{a}_{i,k-1}^*)$.

\subsection*{Appendix 2: Further details on Add Health variable definitions}

Variables used from wave I included: current cigarette smoking, age, gender, race, ethnicity, height, weight, education, exercise, self-rated health, ever tried cigarettes, and alcohol use. Age was determined from self-reported birth year and year of wave I interview. Gender (male, female) was determined by the interviewer, with the option to ask the participant if necessary. Race was defined as self-reported category that best describes them (White, Black, Asian or Pacific Islander, Native American). Ethnicity was self-reported as Hispanic or non-Hispanic. Height and weight were self-reported in inches and pounds, respectively. Education was defined as the current grade level of the participant (7\textsuperscript{th}, 8\textsuperscript{th}, 9\textsuperscript{th}, 10\textsuperscript{th}, 11\textsuperscript{th}, 12\textsuperscript{th}). Exercise (not at all, 1-2 times, 3-4 times, 5+ times) was based on self-reported number of times exercised during the previous week, with provided examples to participants including: jogging, walking, karate, jumping rope, gymnastics, or dancing. Self-rated health was the self-reported general health from the following options: excellent, very good, good, fair, poor. Ever tried cigarettes (yes, no) was defined as self-reporting ever trying cigarette smoking, even just 1 or 2 puffs. Alcohol use was defined as self-reported days drinking alcohol in the previous 12 months (zero or never, 1-2 days in the past 12 months, 1-3 days a month or between 3-12 times in the past 12 months, 1-2 days a week, 3 or more days a week).

Variables used from wave III in this analysis included: current cigarette smoker, age, gender, race, ethnicity, height, weight, education, exercise, self-rated health, alcohol use, previous diagnosis of high blood pressure or hypertension, and current health insurance status. Unless noted otherwise, variable definitions were the same across waves I and III. Our analysis allowed self-reported gender, race, and ethnicity to vary between waves I and III, as a participants' self-identification can vary over time. Measures for gender, education, height, weight, and exercise differed between waves I and III. Gender was self-reported (male, female). Education was defined as highest completed education (less than high school, high school, at least some college, pursuit of graduate degree). Height and weight were instead measured at wave III by the interviewer. As the scales used only measured up to 330, those with a weight above 330 were considered missing. For the definition of exercise, a single exercise variable was constructed from the following questions:
\begin{itemize}
	\item ``In the past seven days, how many times did you bicycle, skateboard, dance, hike, hunt, or do yard work?"
	\item ``In the past seven days, how many times did you roller blade, roller skate, downhill ski, snow board, play racquet sports, or do aerobics?"
	\item ``In the past seven days, how many times did you participate in strenuous team sports such as football, soccer, basketball, lacrosse, rugby, field hockey, or ice hockey?"
	\item ``In the past seven days, how many times did you participate in individual sports such as running, wrestling, swimming, cross-country skiing, cycle racing, or martial arts?"
	\item ``In the past seven days, how many times did you participate in gymnastics, weightlifting, or strength training?"
	\item ``In the past seven days, how many times did you play golf, go fishing or bowling, or play softball or baseball?"
	\item ``In the past seven days, how many times did you walk for exercise?"
\end{itemize}
For each question, participants reported the number of times the activity was done, ranging from 0 to 7+. To construct a similar definition of exercise between waves, the reported times for each activity were summed together and then categorized as 0, 1-2, 3-4, and 5+. Previous diagnosis of high blood pressure or hypertension (yes, no) was self-reported. Current health insurance status (yes, no or unknown) was defined by self-reporting health insurance coverage by one of the following: parents', husband's/wife's, work, union, school, military, private, Medicaid, or Indian Health Service.

\end{document}